\documentclass[authoryear]{elsarticle}

\usepackage{natbib}

 \usepackage{graphicx}
\usepackage{amssymb}

\journal{New Astronomy}

\begin{document}

\begin{frontmatter}



\title{On the apsidal motion of BP~Vulpeculae}


\author[auth1]{Csizmadia, Sz.}
\ead{szilard.csizmadia@dlr.de}
\author[auth2]{Ill\'es-Alm\'ar, E.}
\author[auth3]{Borkovits, T.}

\address[auth1]{Institute of Planetary Research, German Aerospace Center,
D-12489 Berlin, Rutherfordstrasse 2., Germany}

\address[auth2]{Konkoly Observatory, H-1525 Budapest, P. O. Box 67., Hungary}

\address[auth3]{Baja Astronomical Observatory, H-6500 Baja, Szegedi \'ut Kt.
766, Hungary}

\begin{abstract}
BP~Vulpeculae is a bright eclipsing binary system showing apsidal motion. It
was found in an earlier study that it shows retrograde apsidal motion which
contradicts theory. In this paper we present the first $BV$ light
curve of the system and its light curve solution as well as seven new times of
the minima from the years 1959-1963. This way we could expanded the baseline of
the investigation to five decades. Based on this longer baseline we concluded
that the apsidal motion is prograde agreeing with the theoretical expectations
and its period is about 365 years and the determined internal  structure
constant is close to the theoretically expected one.
\end{abstract}


\end{frontmatter}

% main text
\section{Introduction}
%\label{Introduction}

\begin{figure}
\begin{center}
\includegraphics[width=12cm]{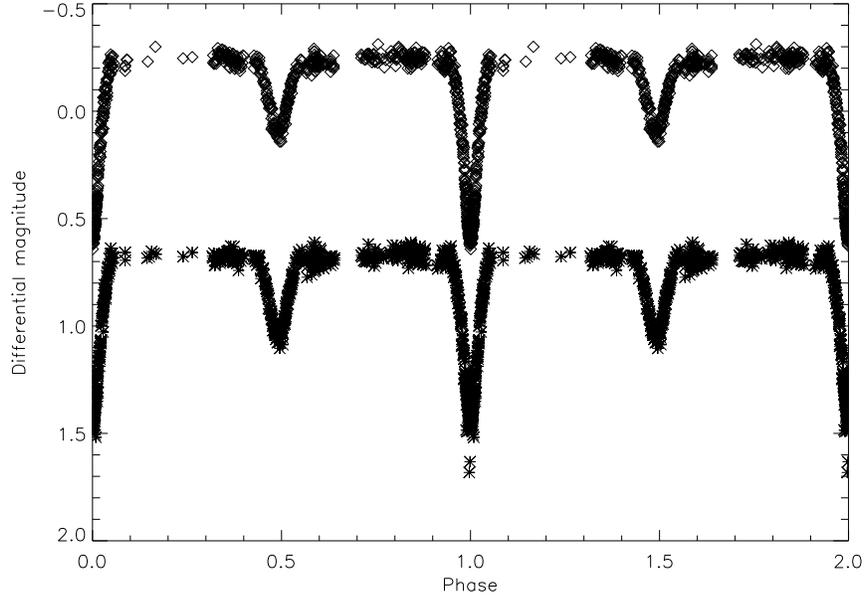}
\end{center}
\caption{The $B$ (top) and $V$ (bottom) light curves of BP~Vul obtained 
in the years 1959-1963. Bottom is the B curve while top curve is the V 
one which is shifted by 0.3 magnitudes for the sake of clarity.}
\label{fig:lightcurve}
\end{figure}

The eclipsing nature of the $10^{th}$ magnitude star BP Vulpeculae was discovered by
\citet{illesalmar60} and she gave a period value of $1.938$ days which was
slightly corrected by \citet{huth65}. Then the star was neglected until the end of
the 20th century when \citet{lacy92} published $UBV$ colours of the system at
certain phases. {Later \citet{lacyetal03} presented more than 5000 data points in one
colour ($V$) and they have solved their light curve and determined very
accurately absolute dimensions of the system. This $V$ light curve was again
solved by a simple, but automatized code \citep{devor05} and the result was
similar to that of \citet{lacyetal03}.

The system has an age of about 1 Gyr consisting of A7m V + F2m V
spectral type components and is slightly eccentric ($e=0.0345$,
\citealp{lacyetal03}). In eccentric binary star systems the apsidal line
is revolving which is called apsidal motion and therefore the time
difference between a primary minimum and the subsequent secondary
minimum is variable. The apsidal motion is usually caused by two
effects: the tidal forces of the components and the effects of general
relativity. In the case of a more or less ordinary eccentric binary,
i.e. no very close third companion in the system, and no extreme stellar
rotation, both phenomena predict that the apside shows a prograde
motion. The total apsidal motion is a sum of the classical and
relativistic contributions. It is worthy to mention that there are
several systems, like DI~Herculis or AS~Camelopardalis where the
observed apsidal motions highly differ from the theoretically predicted
ones \citep[see e.g.][and references therein]{claret98}. It seems that
the mentioned other effects, like third bodies etc. cause these
pecularities \citep[for an overview see][]{borkovitsetal07}.

Regarding the case of BP~Vul, \citet{lacy03} established a very short apsidal
motion period ($77\pm22$yr) and they found that it is a retrograde one. If
BP~Vul really would have had a retrograde apsidal motion it would be a new
representative of the systems which confronts theory and requires further
study because a retrograde apsidal motion is in contradiction with
theory. But the observational window in the work of \citet{lacy03} was about
one decade only therefore this rough estimation for the period of apsidal motion
should be refined. For this refinement we present $BV$ observations of BP~Vul
which were obtained 45-49 years ago and were not published until now.

\section{Observations \label{Observations}}

The $BV$ data points of BP Vul published here were obtained by one of us (E.
Ill\'es-Alm\'ar) during the years 1959-1963. The observations were carried out
by the 60 cm Newtonian-telescope of the Konkoly Observatory, which is located
at Budapest and it was installed in 1926. The detector was an 1P21 RCA
photoelectric tube and $B$ and $V$ filters were used. The observations were
reduced via a standard way. These differential magnitude data (they had been
measured to a comparison star and transformed into standard system) are
published here (see Tables 1-2 which are available electronically via the
SIMBAD homepage\footnote{http://simbad.u-strasbg.fr/simbad/}).  The $B$ and
$V$ curves can be seen in Figure~\ref{fig:lightcurve}.

Using the method of \citet{kweevanwoerden56} from these observations we
determined seven new times of minima which are reported in Table~3}. All the
minima times we used for the analysis can be found in that Table, too.

\addtocounter{table}{2}

\begin{table}
\begin{center}
\caption{Observed times of minima of BP Vul. Meaning of weights (W): 10: CCD/photoelectric minimum, 2: plate minimum, 1: visual observation, 0: not used for
the calculation because it is an outlier. 
\label{tab:ToM}}
{\small
\begin{tabular}{cccc|cccc}
\hline
\hline
Time of Min. & W & Type & Ref. & Time of Min. & W & Type & Ref. \\
(HJD-2~400~000) &        &      &      & (HJD-2~400~000) &        &      &    \\
\hline
21787.723    &   2  &	p&	1 &37898.427	&   2  &   p&	   2    	       \\
22144.750    &   2  &	p&	1 &37933.365	&   2  &   p&	   2    	       \\
22637.569    &   2  &	p&	1 &37935.294	&   2  &   p&	   2    	       \\
23607.741    &   2  &	p&	1 &38001.256	&   2  &   p&	   2    	       \\
25831.362    &   2  &	p&	2 &38255.437	&   2  &   p&	   2    	       \\
26465.843    &   2  &	p&	1 &38288.425	&  10  &   p&	   3\\
26512.439    &   2  &	p&	2 &38323.326	&   2  &   p&	   2\\
26535.718    &   2  &	p&	1 &38614.433	&   2  &   p&	   2\\
26545.469    &   2  &	p&	2 &38938.446	&   1  &   p&	   4\\
26647.308    &   2  &	s&	2 &38938.452	&   1  &   p&	   4    	       \\
26648.332    &   2  &	p&	2 &38938.453	&   1  &   p&	   4 \\
26868.522    &   2  &	s&	2 &41082.509	&   1  &   p&	   5 			\\
26930.399    &   0  &	p&	2 &41107.549	&   0  &   p&	   5 			\\
27965.784    &   2  &	p&	1 &41115.516	&   1  &   p&	   5 			\\
28074.410    &   2  &	p&	2 &41965.387	&   1  &   p&	   6 			\\
28078.299    &   2  &	p&	2 &42386.456	&   2  &   p&	   1 			\\
28460.527    &   2  &	p&	1 &42666.772	&   2  &   s&	   1 \\
29073.744    &   2  &	p&	1 &43317.780	&   2  &   p&	   1 \\
29114.476    &   2  &	p&	2 &43423.529	&   2  &   s&	   1 \\
29857.637    &   2  &	p&	1 &44757.540	&   1  &   p&	   7  \\
31318.694    &   2  &	p&	1 &44875.875	&   2  &   p&	   1 			\\
33854.692    &   2  &	p&	1 &45114.551	&   1  &   p&	   8 			\\
34209.792    &   2  &	p&	1 &45504.566	&   1  &   p&	   9 			\\
34221.446    &   2  &	p&	2 &45541.452	&   1  &   p&	   10			\\
34580.427    &   2  &	p&	2 &45611.295	&   1  &   p&	   11			\\
35224.585    &   2  &	p&	2 &45611.302	&   1  &   p&	   11\\
35226.576    &   2  &	p&	2 &45636.490	&   2  &   p&	   1 \\
35721.355    &   2  &	p&	2 &45785.882	&   2  &   p&	   1 			\\
36433.446    &   2  &	p&	2 &45855.778	&   2  &   p&	   1 			\\
36790.450    &   2  &	p&	2 &45931.461	&   1  &   p&	   12			\\
36859.3445   &  10  &	s&	3 &45933.404	&   1  &   p&	   13			\\
36860.331    &  10  &	p&	2 &46003.248	&   1  &   p&	   14			\\
36860.3311   &  10  &	p&	3 &46290.424	&   1  &   p&	   15			\\
37116.458    &   2  &	p&	2 &46321.464	&   1  &   p&	   15			  \\
37438.547    &   2  &	p&	2 &46356.387	&   1  &   p&	   15			  \\
37506.4677   &  10  &	p&	3 &46385.508	&   2  &   p&	   1 			  \\
37543.3344   &  10  &	p&	3 &46534.864	&   2  &   p&	   1 			  \\
37572.4389   &  10  &	p&	3 &46612.497	&   1  &   p&	   16			  \\
37642.260    &   2  &	p&	2 &46612.497	&   1  &   p&	   16			  \\
37867.3714   &  10  &	p&	3 &46612.500	&   1  &   p&	   16			  \\
\hline
\end{tabular}}
\end{center}
\small{1: \citet{torresguilbault03}; 2: \citet{huth65}; 3: Present paper;
	   4: BAV 7; 5: BBSAG 30; 6: BBSAG 12; 7: BAV 34; 8: BBSAG 61; BBSAG 67
       10: BAV 38; 11: BBSAG 69; 12: BAAVSS 61; 13: BBSAG 73; 14: BBSAG 74; 15: BBSAG 78; 16: BRNO 28}
\end{table}
%
%
%\addtocounter{table}{-1}
%
%
%\begin{table}
%\begin{center}
%\caption{(Continue.)}
%\begin{tabular}{cccc}
%\hline
%\hline
%Time of Minimum & Weight & Type & Reference \\
%(HJD-2~400~000) &        &      &           \\
%\hline\hline
%
%\hline
%\end{tabular}
%\end{center}
%\end{table}
%

\addtocounter{table}{-1}

\begin{table}
\begin{center}
\caption{(Continue.)}
{\small
\begin{tabular}{cccc|cccc}
\hline
\hline
Time of Min. & W & Type & Ref. & Time of Min. & W & Type & Ref. \\
(HJD-2~400~000) &        &      &      & (HJD-2~400~000) &        &      &    \\
\hline\hline
46612.501    &   1  &	p&	16 &49216.450	&    1&     p&     29			 \\
46612.502    &   1  &	p&	16 &49216.461	&    1&     p&     29			 \\
46612.503    &   1  &	p&	16 &49216.464	&    1&     p&     30			 \\
46612.504   &	 1&	p&     16  &49216.468	&    1&     p&     30			 \\
46612.506   &	 1&	p&     16  &49218.393	&    1&     p&     29			 \\
46612.507   &	 1&	p&     16  &49218.402	&    1&     p&     29			 \\
46612.509   &	 1&	p&     16  &49218.411	&    1&     p&     30			 \\
46614.441   &	 1&	p&     16  &49218.413	&    1&     p&     31			 \\
46614.445   &	 1&	p&     16  &49251.390	&    1&     p&     30 \\
46614.447   &	 1&	p&     16  &49321.242	&    1&     p&     30\\
46614.447   &	 1&	p&     16  &49934.390	&    1&     p&     32		       \\
46614.448   &	 1&	p&     16  &49967.380	&    1&     p&     32	       \\
46614.452   &	 1&	p&     16  &50002.313	&    1&     p&     32		       \\
46678.476   &	 1&	p&     16  &50324.400	&    1&     p&     33		       \\
46678.478   &	 1&	p&     16  &50357.372	&    1&     p&     33		       \\
46709.547   &	 2&	p&     1   &50547.551	&    1&     p&     34 \\  		 
46973.428   &	 1&	p&     17  &50681.418	&    1&     p&     34		       \\
47026.696   &	 2&	s&     1   &50718.283	&    1&     p&     35 \\  		 
47039.370   &	 1&	p&     18  &50751.277	&    1&     p&     35		       \\
47064.618   &	 2&	p&     1   &51036.496	&    1&     p&     36 \\  		 
47361.497   &	 1&	p&     19  &51063.6717  &   10&	  p&	 37	       \\
47363.441   &	 1&	p&     19  &51128.645   &   10&	  s&	 37	       \\
47392.530   &	 1&	p&     20   &51129.646   &   10&	  p&	 37	       \\
47392.535   &	 1&	p&     20   &51327.564   &    1&	  p&	 38		       \\
47392.543   &	 1&	p&     20   &51364.416   &    1&	  p&	 39		       \\
47396.424   &	 1&	p&     19  &51397.4114  &   10&	  p&	 40\\
47431.331   &	 1&	p&     21  &51464.3104  &   10&	  s&	 40\\
47466.271   &	 1&	p&     21  &52031.90450 &   10&	  p&	 41       \\
47788.3674  &	10&	p&     22  &52064.89086 &   10&	  p&	 41       \\
47790.313   &	 1&	p&     23  &52096.86757 &   10&	  s&	 41       \\
47823.313   &	 1&	p&     24  &52098.80834 &   10&	  s&	 41       \\
48112.405   &	 1&	p&     25  &52099.8166  &   10&	  p&	 41       \\
48112.412   &	 1&	p&     25  &52101.75702 &   10&	  p&	 41       \\
48147.327   &	 1&	p&     25  &52164.7794  &   10&	  s&	 41       \\
48176.432   &	 1&	p&     25  &52165.78900 &   10&	  p&	 41       \\
48533.457   &	 1&	p&     26  &52425.79570 &   10&	  p&	 42       \\ 
48723.609   &	 1&	p&     27 &52487.88765 &   10&	  p&	 42       \\
48859.422   &	 1&	p&     28 &52488.81917 &   10&	  s&	 42       \\
\hline
\end{tabular}}
\end{center}
{\small        17: BBSAG 84; 18: BBSAG 86; 19: BBSAG 89; 20: BRNO 30; 21: BBSAG 90; 22: BAV 56
23: BBSAG 92; 24: BBSAG 93; 25: BBSAG 96; 26: BBSAG 99; 27: BBSAG 101; 28: BBSAG 102
29: BRNO 31; 30: BBSAG 105; 31: BBSAG 104; 32: BBSAG 110; 33: BBSAG 113; 34: BBSAG 115; 35: BBSAG 116
36: BBSAG 118; 37: \citet{lacyetal99}; 38: BBSAG 120; 39: BRNO 32; 40 \citet{agereretal01}; 
41: \citet{lacyetal02}; 42: \citet{lacy02}}
\end{table}

\addtocounter{table}{-1}

\begin{table}
\begin{center}
\caption{(Continue.)}
{\small
\begin{tabular}{cccc|cccc}
\hline
\hline
Time of Min. & W & Type & Ref. & Time of Min. & W & Type & Ref. \\
(HJD-2~400~000) &        &      &      & (HJD-2~400~000) &        &      &    \\
\hline\hline
52495.64880 &	10&	p&     42  &53526.9042  &   10&     s&     47  \\
52562.5517  &	10&	s&     42  &53527.91289 &   10&     p&     47  \\
52595.5379  &	10&	s&     42  &53898.5192  &   10&     p&     48  \\
52724.589   &    1&     p&     43  &53933.4432  &   10&     p&     49  \\
52782.8192  &   10&     p&     44  &53987.7740  &   10&     p&     50  \\
52814.7964  &   10&     s&     44  &54026.5809  &   10&     p&     50  \\
52817.74512 &   10&     p&     44  &54325.3939  &   10&     p&     51  \\
53169.8789  &	10&	s&     45  &54388.4173  &   10&     s&     51  \\
53186.4111  &	10&	p&     46  &            &     &      &         \\
\hline
\end{tabular}}
\end{center}
{\small 43: \citet{diethelm03}; 44: \citet{lacy03}; 45: \citet{lacy04}; 46: \citet{zejda04}; 47: \citet{lacy06};
48: \citet{hubscheretal06}; 49: \citet{hubscher07}; 50: \citet{lacy07}; 51: \citet{hubscheretal08}}
\end{table}

\section{Period analysis}												 
\label{Introduction}

\begin{figure}
\begin{center}
\includegraphics[width=12cm]{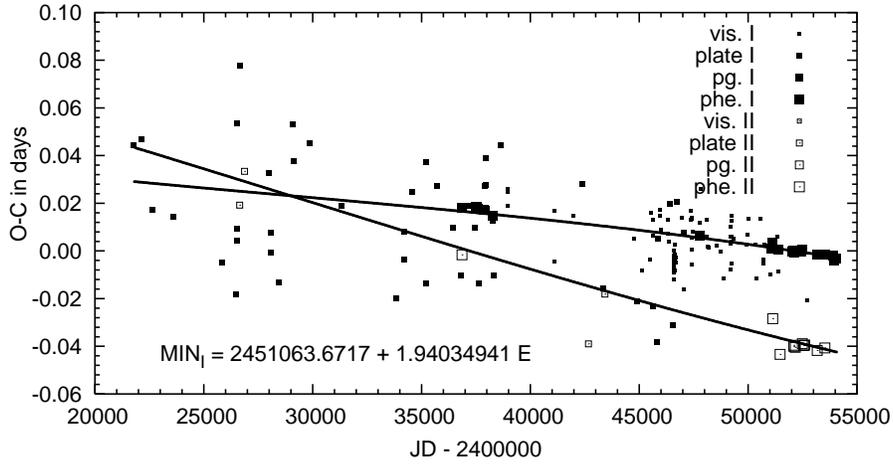}
\end{center}
\caption{The $O-C$ diagram of BP Vulpeculae. Filled and open squares represent primary and secondary minima, respectively. Lines show the fits to the $O-C$ residuals for the primary and secondary minima, respectively. The weights of the different kind of minima can be found in the text.}
\label{fig:O-C}
\end{figure}

The $O-C$ values were calculated with the following ephemeris:
\begin{equation}
\mathrm{Min~I} = 2\,436\,860.3311 + 1.9403494 \times E
\end{equation}
where the initial epoch was our first primary minimum observation while
the period was taken from \citet{lacyetal03}. The $O-C$ diagram is
presented in Figure~\ref{fig:O-C}. It is clear from Figure~\ref{fig:O-C} that the time lag between
the primary and secondary minima has changed so the apsidal motion is
clearly present. In the following calculations CCD and photoelectric times
of minima had weights of 10, plate minima had 2, visual observations had
1.

Using a second-order approximation in the eccentricity, the times of
primary and secondary minima will occur at the times given below:
\begin{equation}
{\mathrm{Min~I}} = T_0 + EP_{\mathrm{s}} - \frac{eP_{\mathrm{a}}}{\pi}\cos\omega_E +
\frac{3}{8}\frac{e^2 P_{\mathrm{a}}}{\pi} \sin 2\omega_E  + ...
\end{equation}
\begin{equation}
{\mathrm{Min~II}} = T_0 + EP_{\mathrm{s}} - \frac{P_{\mathrm{a}}}{2} + \frac{eP_{\mathrm{a}}}{\pi}\cos\omega_E +
\frac{3}{8}\frac{e^2 P_{\mathrm{a}}}{\pi} \sin 2\omega_E  - ...
\end{equation}
where $T_0$ are the epoch of a primary minimum, $P_{\mathrm{a}}$ is the
anomalistic period, $P_{\mathrm{s}}$ is the sidereal period, i.e.
$P_{\mathrm{s}} \approx P_{\mathrm{a}} (1 - \omega' / 2\pi)$, $e$ is the
eccentricity, $\omega_E = \omega_0 + \omega' E$ where $\omega' = 2\pi
P/U$. $U$ is the apsidal motion period.

Applying these formulae we determined the apsidal motion period with the
upgraded version of LiteAM software developed by T. Borkovits \citep[see e.g.][]{borkovitsetal02}. We found from the fitting of the $O-C$ curve that
$U/P = 68700 \pm 500$ and this means $U=365$ years, $T_0 = 51063.6537
\pm 0.0001\mathrm(HJD)$ and $\omega_0 = 150^\circ \pm 5^\circ$. This
latter value is in good agreement with $\omega=154.7^\circ \pm 3.9^\circ$
found by spectroscopic measurements \citep{lacyetal03}. The apsidal motion
period yields $\dot \omega \approx 1.0^\circ$/year.

Note that the $O-C$ curve is not well-covered yet. There is need for more
observations to determine an exact value of the apsidal motion period in
BP~Vul -- our value given above can be regarded as a first approximation.
But more interesting than the exact value of this period is that the
new value yielded a prograde motion of the semi-major axis instead of
a retrograde one.

\section{Light curve solution}												 
\label{LCSolution}													 

For the light curve solution we used the Wilson-Devinney Code \citep{wilson98}. 
The free parameters were the inclination, the dimensionless
surface potentials, argument of periastron and its time-derivative and
the luminosities of the components. Limb-darkeking coefficients were
fixed and these fixed values were interpolated ones from tables of \citet{vanhamme93}. 
Gravity darkening and reflexion coefficients were also
fixed. Mass ratio, surface temperatures of the components and
eccentricity of the orbit were fixed at the values given in \citet{lacyetal03}. 
Then differential correction analysis were carried out and the
stopping criteria was that the change in the parameters in the final
step should be lower than its standard deviations. Since BP Vul has a
fast apsidal motion (see previous Section) we used time as an
independent variable during the modeling rather than phase. The result
of the light curve solution can be found in Table~\ref{table:LC}.

Comparing our results to the one of \citet{lacyetal03} we found a
remarkably excellent agreement in luminosity ratio, but other elements
are slightly different. However, the precision of our light curve does
not reach the precision of their one although we have colour
information, too. Moreover, they used the so-called EBOP code \citep{popper81} 
which has a slightly different input physics. Since we solve these
old light curves for the purpose to determine the argument of periastron
independently, these slight differences do not destroy the validity
of our light curve solution.

Thus we concentrate the position of the periastron hereafter. As one can
see from Table~4 $\omega = 126^\circ \pm 5^\circ$ at epoch HJD
=$2\,436\,860.3311$ (our adopted epoch) which nearly corresponds 1959
October 18. From their spectroscopic measurements \citet{lacyetal03} had
given $\omega = 154.7^\circ$ for epoch HJD $2\,451\,023.254$ which
nearly corresponds to 1998 July 28. It is easy to compute that these two
measurements yield $0.74^\circ$/yr apsidal motion or $U=486\pm57$ years.

This value is more than the 365 years apsidal motion period --
determined from the $O-C$ analysis -- by about 120 years. According to
us this 33\% difference is not because the $O-C$ diagram is not
well-covered and a new light curve solution based on multi-colour
observations are needed.

However, the fitted $\dot \omega$ gives $1.0^\circ$/yr which is fully 
in agreement with the results of the $O-C$ analysis. Regarding the 
uncertainty in the position of $\omega$ determined from these old 
light curves, one can conclude that very likely the light 
curve solution gives $\dot \omega \approx 1.0^\circ$/yr.

\begin{table}
\label{table:LC}
\begin{center}
\caption{Light curve solution of BP~Vulpeculae. Denotions have their 
usual meaning. Mode~0 of the Wilson-Devinney Code was used 
\citep[see][]{wilson98}. $r_{1,2}$ were derived from $\Omega_{1,2}$ and 
$q$ by the WD-code itself.}
\begin{tabular}{cccc}
\hline
\hline
Quantity             &          & This paper     & \citet{lacyetal03} \\
\hline
$i$                  & adjusted & $86.64\pm0.16$ & $87.67$ \\
$L_1 / L_\mathrm{tot}$ (B) & adjusted & $0.719\pm0.02$ & - \\
$L_1 / L_\mathrm{tot}$ (V) & adjusted & $0.696\pm0.02$ & $0.718$ \\
$\omega$             & adjusted & $126^\circ\pm5^\circ$ & $154.7^\circ$ \\
$\dot \omega$        & adjusted & $1.006^\circ\pm0.002^\circ$ & - \\
$\Omega_1$           & adjusted & $7.04\pm0.09$  & - \\
$\Omega_2$           & adjusted & $6.89\pm0.07$  & - \\
$HJD0$               & fixed    & 2 436 860.3311 & 2 451 023.254 \\
mean $r_{1}$   & derived& $0.162\pm0.027$ & 0.1931 \\
mean $r_{2}$   & derived& $0.141\pm0.031$ & 0.1552 \\
$g_1$                & fixed       & $1.0$   & -             \\
$g_2$                & fixed       & $1.0$   & -             \\
$A_1$                & fixed       & $1.0$   & -             \\
$A_2$                & fixed       & $1.0$   & -             \\
$T_1$                & fixed       & $7709$K & $7709$K       \\
$T_2$                & fixed       & $6823$K & $6823$K       \\
$x_{1, \mathrm{bol}}$ & fixed       & $0.538$ & -             \\
$x_{2, \mathrm{bol}}$ & fixed       & $0.467$ & -             \\
$x_{1, \mathrm{B}}$   & fixed       & $0.604$ & -             \\
$x_{2, \mathrm{B}}$   & fixed       & $0.621$ & -             \\
$x_{1, \mathrm{V}}$   & fixed       & $0.534$ & $0.50\pm0.03$ \\
$x_{2, \mathrm{V}}$   & fixed       & $0.507$ & $0.56\pm0.03$ \\
$e$                  & fixed       & $0.0345$& $0.0345$      \\
$q$                  & fixed       & $0.811$ & $0.811$       \\

\hline\hline
\hline
\end{tabular}
\end{center}
\end{table}

\section{The internal structure constant $k_2$}
\label{k2}

Our next calculations are based on \citet{gimenez85}. From the known
eccentricity, masses and period given in \citet{lacyetal03} one can
calculate that the relativistic contribution to the apsidal motion in
BP Vulpeculae is $7.529 \cdot 10^{-4}$ degree/cycle. From the observed
$U=365$ years see above we found $\dot \omega_\mathrm{obs} = 5.24 \cdot
10^{-3}$ degree/cycle. So the Newtonian term in the apsidal motion is
$\dot \omega_N = \dot \omega_\mathrm{obs} - \dot \omega_\mathrm{rel} = 
4.49 \cdot 10^{-4}$ degree/cycle.

Using the well-known relationship
\begin{equation}
k_{2, \mathrm{obs}} = \frac{1}{c_{21} + c_{22}} \frac{\dot \omega_N}{360^\circ}
\end{equation}
we found $\log k_{2, \mathrm{obs}} = -2.66\pm0.08$. Here $c_{21}$ and $c_{22}$
are the functions of eccentricity, mass ratio and fractional radii of the
components and their precise form is given in Gimenez (1985).

Using the tables of \citet{claret04} with X=0.70, Z=0.02, t=1 Gyr and with
mixing length parameter $\alpha=1.68$ and overshooting parameter
$\alpha_{OV}=0.2$ we could calculate $\log k_{2, \mathrm{theo}} =
-2.47$. Regarding the uncertainties in the determined apsidal motion
period which appears in the determination of $k_2$ internal structure
constant we may conclude that the observed and theoretically expected
values are close to each other. Also note that we have only one
secondary minimum from the 1960s which is a key point in similar
calculations.

\section{Summary}
\label{Summary}

BP~Vulpeculae is an eccentric eclipsing binary star showing the
so-called periastron-precession effect. \citet{lacyetal03} concluded
that this effect causes a retrograde motion of the semi-major axis and
it has a period of $77\pm22$ years based on their about one decade
long observational material. However, retrograde motion contradicts
theory. Their explanation was that a possible third body in the system
could perturb the orbit yielding the observed peculiar periastron
precession. Nevertheless no spectroscopic evidence was found by them
for such a third body.

BP~Vul was observed at the Konkoly Observatory more than forty years
before the work of \citet{lacyetal03} by one of the authors of this study.
This observational material allowed us to determine the times of six
primary and one secondary minima. With these early observations the
baseline could be expanded to approximately five decades which was
enough to refine the apsidal motion period determined by \citet{lacyetal03}. 
Our $O-C$ analysis based on the extended time-line showed that
an unseen, dark third body in the system cannot be extracted from the
presently available minima observations. All of these makes very
unlikely the presence of a third body with a mass and orbit which
would cause a peculiar periastron precession.

First time two-colour light curves were presented by us for BP~Vul.  The
light curve solution -- using the Wilson-Devinney Code -- yielded very
similar results comparing to \citet{devor05}'s one and
\citet{lacyetal03}'s one . In addition, the $O-C$ analysis in this paper
showed that the apsidal motion period in BP~Vul is prograde and it has a
period of about 365 years -- it is in agreement with the value
determined with less accuracy from the light curve. The prograde motion
means that BP Vul is not a representative of problematic cases and it is
in agreement with theoretical expectations. Nevertheless we concluded
that the negative apsidal motion rate determined by \citet{lacyetal03}
is only a consequence of their short observational window. 

We also calculated the $k_2$ internal structure of BP~Vul and found it
being close to the theoretical value. The slight difference should be
refined in the future with a better observed $O-C$ diagram without
gaps. Therefore the minima observations of BP~Vulpeculae in the future
are needed.

The comments on the first version of the manuscript by Drs J. Jurcsik and K.
Ol\'ah is acknowledged.

\end{document}